\documentstyle[11pt,aaspp4]{article}

\lefthead{Crampton et al.}
\righthead{The SMC Supersoft Binary SMC 13}

\begin{document}

\title{THE SMC SUPERSOFT X-RAY BINARY 1E 0035.4$-$7230 (SMC 13)}

\author{David Crampton\altaffilmark{1} and J.B.
Hutchings\altaffilmark{1}} \affil{Dominion Astrophysical Observatory,
National Research Council of Canada,\\ Victoria, B.C. V8X 4M6, Canada} 

\author{A.P. Cowley\altaffilmark{1} and P.C. Schmidtke\altaffilmark{1} }
\affil{Physics \& Astronomy Dept., Arizona State University, Tempe, AZ, 
85287-1504} 

\altaffiltext{1} {Visiting Astronomers, Cerro Tololo Inter-American
Observatory, National Optical Astronomy Observatories, which is operated
by the Association of Universities for Research in Astronomy, Inc., under
contract with the National Science Foundation}

\begin{abstract}

Details of simultaneous photometric and spectroscopic observations of
the optical counterpart of the ``supersoft" X-ray source in the Small
Magellanic Cloud 1E 0035.4$-$7230 (`SMC 13') are presented.  Although
the spectrum is dominated by emission lines of He II, the Balmer series
of hydrogen is also present in emission with a strong decrement, and 
Balmer lines are seen as broad absorptions.  Several high ionization
emission features are also present including O VI (3811, 3834, and
5290\AA).  Radial velocities and photometry confirm that the binary
period is $\sim$0\fd1719, and an improved value of the period is
derived from four years of photometry and analysis of $ROSAT$-HRI X-ray
data.  The orbital light variation is primarily due to an eclipse of
the extensive accretion disk.  X-ray and optical minima occur
together.  The $UBV$ light curves are similar to each other, and no
clear phase-related color variations are found.  He II emission-line
velocities show a semi-amplitude of K$\sim$100 km s$^{-1}$, and maximum
velocity occurs when the light curve indicates the compact star would
be moving away from the observer, suggesting this emitting region may
trace the orbital motion of the compact star.  The range of possible
masses implied for the X-ray source lies between 0.5 and 1.5M$_{\odot}$
if the mass donor is a main sequence star filling its Roche lobe.  The
light curve suggests values at the high end of this range.  The broad H
absorption lines appear to have a much larger velocity amplitude and
lower systemic velocity, making it difficult to understand their
origin.  We discuss possible models for the system.

\end{abstract} 


\section{INTRODUCTION}

``Supersoft sources'' (SSS) are very luminous X-ray objects (bolometric
luminosities of $\sim$10$^{38}$ ergs s$^{-1}$) with characteristic
blackbody temperatures of kT $\sim$ 30 -- 60 eV (see review by Hasinger
1996).  It now appears that the group contains several different types of
systems including the nuclei of planetary nebulae, symbiotic stars,
compact binaries, and hot white dwarfs.  Even within the group of compact
binaries several different models are possible, but all require accretion
near, or even above, the Eddington limit.  The high X-ray luminosities and
optically bright accretion disks suggest very high mass transfer/accretion
rates, well above that in classical low-mass X-ray binary systems (LMXB)
like Sco X-1 or LMC X-2.  Broad emission and P Cygni absorption profiles
observed in the spectra of several SSS are additional evidence of mass
transfer or loss.  Whether the compact objects in these binaries are white
dwarfs, neutron stars or black holes is still being debated.  The most
widely accepted SSS model involves steady nuclear burning on the surface
of an accreting white dwarf (e.g., van den Heuvel et al.\ 1992, Pakull et
al.\ 1993).  However, the few dynamical studies which have been undertaken
for these binary systems suggest a range of masses for the compact
objects, indicating they may not all be white dwarfs (e.g., Beuermann et
al.\ 1995, Cowley et al.\ 1990, Crampton et al.\ 1996).  If some of the
compact objects are neutron stars or black holes, the soft X-ray spectra
indicate the stars must be surrounded by some type of cocoon which
down-scatters the X-rays (Greiner, Hasinger, \& Kahabka 1991, Kylafis \&
Xilouris 1993, Kylafis 1996). 

Evidence of collimated outflows or ``jets" has been found in two SSS, CAL
83 and RX J0513.9$-$6951 (Crampton et al.\ 1996).  Livio (1996) argues
that since jet velocities are typically equal to the escape velocity from
the central object, the observed SSS jet velocities ($\sim$4000 km
s$^{-1}$ in RX J0513.9$-$6951) indicate that the central objects have
masses appropriate to white dwarfs.  However, analysis of the behavior of
the X-rays and visible light during a brief ``off-state'' of CAL 83 in
1996 suggests that the white dwarf must be massive, very near to the
Chandrasekhar limit (Alcock et al.\ 1997), if the steady nuclear burning
model is correct.  Thus, there is need for more and improved measurements
of the stellar masses in supersoft X-ray binaries.  We are in the process
of obtaining data for several SSS, and this paper reports a new study of
SMC 13. 

\subsection{1E 0035.4$-$7230, SMC 13}

The ``supersoft'' X-ray source 1E 0035.4$-$7230 was discovered during the
Einstein Observatory X-ray survey of the Small Magellanic Cloud (Seward \&
Mitchell 1981).  The source is alternatively known as `SMC 13' (Wang \& Wu
1992, Schmidtke et al.\ 1996), and for simplicity we will use that
designation in this paper.  Its optical counterpart was identified
relatively recently with a faint ($V\sim$20.4) blue star (Orio et al.\
1994, Schmidtke et al.\ 1994).  Schmidtke et al.\ (1996) and van Teesling
et al.\ (1996) discussed early photometric and spectroscopic observations
of this optical counterpart, but neither investigation resulted in a clear
understanding of the nature and masses of the stellar components of the
binary system.  The short period of the system (P$\sim$0.1719 d) has also
hampered the analyses since the relative phasing of data taken during
different observing seasons could not be established, and hence there was
considerable uncertainty in the precise period.  In particular, since the
pointed ROSAT-PSPC X-ray observations (Kahabka 1996) were made between 1
and 2.5 years earlier than the optical observations reported by Schmidtke
et al.\ (1996), the relative phasing and even the shape of the X-ray light
curve could not be determined reliably.  Finally, the very short period
made it difficult to obtain phase-resolved optical spectra. 

In 1996 November, we were able to obtain additional photometry and greatly
improved spectroscopic data for SMC 13 which are reported here and are
combined with our earlier observations.

\section{OBSERVATIONS AND MEASUREMENTS}

\subsection{Photometry}

New photometric observations of SMC 13 were taken with the Tek2048\#2
CCD on the CTIO 0.9-m telescope during five nights in 1995 November and
six nights in 1996 November.  This photometry is presented in Table 1. 
In addition, $B$ and $U$ photometry from 1993 and 1994, also obtained
with the CTIO 0.9-m, which have not been previously published are
presented in the same table.  Our analysis uses our new photometry as
well as the 1993 and 1994 data published by Schmidtke et al.\ (1996). 

The observations were calibrated using Landolt (1992) standard stars and
reduced using DAOPHOT (Stetson 1987).  Differential magnitudes were
calculated for the $B$ and $V$ filters relative to local photometric
standards within the CCD frames using a procedure which minimizes errors
by PSF fitting (Schmidtke 1988).  Since the comparison stars used for the
$B$ and $V$ frames are very red, they proved unsuitable for reducing $U$
filter images.  Hence, a separate set of local standards, considerably
brighter in the ultraviolet, was defined for these data.  The mean errors
are approximately $\pm$0.03 mag, as shown in Table 1.  Overall, the system
is very blue, with mean magnitude and colors of $V=20.4$, $B-V=-0.13$ and
$U-B=-1.13$.

\subsubsection{Photometric Period and Ephemeris}

The period of SMC 13 had previously been found to be near 0.1719 days
(Schmidtke et al.\ 1996), but this short period combined with a separation
of a year between each observing run made it impossible to determine the
exact value, since cycle count was lost.  This meant, for example, that
one could not be certain of how the optical photometry phased with the
1992--93 $ROSAT$ X-ray observations. 

In the present study we have used all of the 1993 -- 1996 $V$ photometry
to search for periods using the routine described by Horne \& Baliunas
(1986).  The data are dominated by the large amount of 1994 $V$
photometry.  The resulting periodogram is displayed in Figure~\ref{per}.
There are three strong peaks corresponding to periods of 0.172007,
0.171925, and 0.171844 days.  Light curves plotted on these periods are
very similar.  Thus, to aid in distinguishing between these possible
periods we carefully examined the data from each run as a whole and those
taken on individual nights (i.e., during a single orbital cycle) to check
for changes in mean level or other peculiarities in the light curves. 

There are clearly variations in both the shape and mean level of the light
curve at different epochs.  As in the supersoft binary CAL 87, the ingress
portion of the `eclipse' shows the most variation.  In CAL 87, and by
analogy perhaps in SMC 13, the changes in this part of the light curve are
attributed to variations in the structure of the outer disk which
partially occults the central bright source.  Because much of the
photometry was obtained in 1994, the light curves for each observing run
and individual nights within a run were compared with the 1994 $V$ light
curve.  The 1993 data should be given lowest weight because a different,
smaller format CCD was on the telescope, and fewer comparison stars were
used in the reduction.  The 1993 and 1995 data show the same mean
magnitude as in 1994, but there is some evidence that the amplitude may be
somewhat smaller in 1995.  However, in the 1995 data there are only a few
points near the expected maximum in the curve so the range is not very
well defined.  On the second night of our 1996 observing run, SMC 13 was
clearly about 0.10 mag brighter than the 1994 average curve, especially
between phases 0.7 and 0.0.  When those points are removed, the remaining
1996 photometry suggests that the system may have been $\sim$0.05 mag
brighter than in 1994.  We have removed all of the 1996 second-night data
from the subsequent period analysis, although they are plotted in the
light curves shown in Figure~\ref{lc}. 

A second method of period determination was then used to examine the
dispersion in phase-binned data and select periods which minimize this
dispersion.  This analysis shows the best periods to be near 0.1719 and
0.1720 days.  Both of these values correspond to the strongest peaks in
the periodogram described above. 

We have also examined all of the $ROSAT$-HRI data which are available in
the public archives.  There is an 800-sec observation taken in 1994 June
and a series of observations taken within two consecutive days in 1995
May.  It is very fortunate that these dates fall within the time covered
by our optical data, so that there is no need to extrapolate back to the
X-ray epoch with an uncertain period.  The 1994 X-ray data were divided
into two 400-sec blocks, and the 1995 data were broken into twelve
$\sim$600-sec pieces.  These data were then phased on both the 0.1719 and
0.1720 day periods.  There is a clear modulation of the X-ray count rate
with a range of about a factor of two when plotted on either of these
periods.  However, using the longer period, the X-ray minimum falls at
optical phase 0.5, while the X-ray minimum occurs at phase 0.0 when the
0.1719 d period is used.  Since virtually all models predict that the X-ray
and optical minima should coincide, at least approximately, and since the
X-ray observations fall about a half-year between our optical data sets,
they allow us to break the uncertainly about the number of cycles between
our optical observations taken one year apart.  Hence we conclude that the
true orbital period of SMC 13 is the shorter of the two ``best" periods, P
$=$ 0.171925$\pm$0.000001 days.  Although the period finding routines
allow us formally to derive a period with slightly higher precision,
trials with removing data from different nights or portions of a night
show the accuracy in determining the period is limited to the errors given
here.  The $ROSAT$-HRI data are also plotted in Figure~\ref{lc}.

The determination of the time of minimum light, T$_0$, is not
straightforward, partly because the shape of the optical ingress and
egress change from observing season to observing season.  In addition,
there is considerable scatter in the light curve both from intrinsic
flickering and from CCD photon noise due to the faintness of the system.
We have adopted a value of T$_0$ which is the mean of the lowest points in
the 1994 $V$ light curve, giving an ephemeris of: 

\medskip \centerline{T$_0 =$ HJD 2,449,664.591 $\pm$ 0.003 $+$ 0.171925E 
$\pm$ 0.000001 days} 

\smallskip
\noindent
The error in T$_0$ is our best estimate of the uncertainty in determining
the time of minimum light.  Since the optical photometry and spectroscopy
were obtained on the same nights, any uncertainty in the precise period
does not affect the comparison of the phase-related variations discussed
below. 

The older $ROSAT$-PSPC data which are given by Kahabka (1996) were
obtained over a timescale of $\sim$1.5 years with only one or two points
being taken on any single day.  This means that it is nearly impossible to
determine the orbital period from these data, since changes in the X-ray
flux level on timescales of days or weeks (as seen in the optical data)
confuse any orbital modulations.  Furthermore, because the data were taken
in scan mode for the $ROSAT$ All Sky Survey (RASS), it was necessary to
make aspect corrections, so that the published count rates contain that
further uncertainty.  As was shown by Schmidtke et al.\ (1996) and Kahabka
(1996), the long-term PSPC X-ray data show some modulation on the orbital
period, but the time of minimum is largely dominated by three low points.
Their occurrence before phase zero suggests they may be caused by the
system being in a low state rather than by the orbital variation seen in
the HRI data. 

\subsubsection{Light Curve, Colors, and X-ray Variability of SMC 13}

The $V$, $B$, and $U$ light curves, based on the ephemeris given above,
are plotted in Figure~\ref{lc}.  $B-V$ colors have been computed using
only those values of $B$ that were bracketed in time by two $V$
observations, or vice versa.  The mean magnitude of the appropriate
bracketing pair was used in calculating a $B-V$ color.  Not all of the
data could be used with this technique, but this method gives the most
reliable colors for this continuously varying binary system.  The $B-V$
color is plotted versus phase in Figure~\ref{lc} also.  There is no
evidence for a systematic variation with phase.  Since the system is so
faint, very little $U$ photometry was obtained, and unfortunately phase
zero is not well covered.  It appears that the magnitude range is similar
in all three colors ($\Delta$m$\sim0.3$ mag).  However, we note that most
of $U$ points during phases 0.6 to 0.9 are lower than would be expected
from the $V$ and $B$ light curves.  For comparison, in CAL 87 the largest
variations in the light curves are also seen in this phase range,
suggesting there may be changes in the disk structure which causes the
occultation.  In SMC 13 such changes may sometimes cause the hottest part
of the disk to suffer greater obscuration, hence depressing the $U$ light
more than the other colors in this part of the orbit.  However, we have
insufficient data to establish a clear phase-related color variation. 

We note that the optical amplitude in SMC 13 is relatively small
($\sim$0.3 mag) while the X-ray count rate apparently varies by about a
factor of two.  By contrast, in the high inclination, optically eclipsing
SSS CAL 87 ($\Delta$m$_V\sim1.2$ mag) the X-ray variation is less than a
factor of two (Schmidtke et al.\ 1995).  This points out that a range of
conditions probably exists among the various supersoft binary systems.
Below we discuss the interpretation of the light curve and possible models
for the system.

\subsection{Spectroscopy}

\subsubsection{Data}

The new spectroscopic data were obtained with the CTIO 4-m telescope
during five nights in November 1996, with the KPGL1 grating and Loral 3K
detector.  The spectra cover the wavelength range $\sim$3700--6700\AA\ and
have a resolution of $\sim$1.0\AA\ per pixel.  With a 1\farcs5 slit,
corresponding to three pixels, the spectral resolution is $\sim$3\AA.  Due
to the faintness of the star and its short period, the length of the
exposures (2400 s, corresponding to 0.16P in phase) was a compromise
between achieving reasonable signal-to-noise and minimizing the duration
of orbital phase covered by each integration.  A total of 17 spectra of
SMC 13 were obtained.  One-dimensional spectra were extracted and processed
following standard {\sc IRAF} techniques to yield wavelength-calibrated
spectra which have a peak S/N $\sim$12.  Calibration spectra (He-Ar) were
taken before and after each stellar exposure, and the wavelengths are
established to $\sim$0.1 pixel or better.  Details of the measurements
made from these spectra are given in Tables 2 and 3.

\subsubsection{The Spectrum of SMC 13}

The spectrum of SMC 13 is characterized by moderately strong emission
lines of He II and hydrogen, particularly He II-4686\AA\ and H$\alpha$, on
a very blue continuum (see Figure~\ref{spec}).  The strengths of those He
II Pickering lines which lie between the Balmer lines (e.g., 5411, 4541,
4200\AA, etc.) show that only at H$\alpha$ does hydrogen contribute
significantly to the blended H$+$He II emission lines near the Balmer
wavelengths.  Higher He II Pickering lines, not normally seen in spectra
of X-ray binaries, are found when the spectra are co-added.  O~VI emission
is also present at 3811, 3835, and 5290\AA.  Several weak unidentified
emission features which are seen in other SSS appear to be present in SMC
13, including lines near 4495\AA\ and 6380\AA.  These may be possibly be N
IV lines, but we note that there is no evidence of N IV at 4057\AA.  Also,
neither C III nor N III emissions in the 4630--4650\AA\ range are present.
The unidentified line near 6380\AA\ might be [Fe X] 6374\AA\ which is seen
in some symbiotic stars and recurrent novae.  There is no evidence of the
C IV lines (5801 and 5812\AA) which are present in some other supersoft
systems (e.g. CAL 83 and RX J0513.9$-$6951).  The emission at 4200\AA\ is
partly an instrumental artifact and not an anomalously strong He II line. 

SMC 13 spectra, co-added in phase bins, also reveal broad absorption
features at the Balmer lines which appear on both sides of the emission
cores (see Figure~\ref{coad}).  Both emission and absorption are visible
together at H$\beta$; at the higher Balmer lines the steep emission
decrement leaves only the broad absorption line, while at H$\alpha$ the
emission feature dominates.  When the spectra are co-added in the
restframe of H$\gamma$ absorption, the equivalent widths of H$\delta$,
H$\gamma$ and H$\beta$ absorption can be measured; they are 0.5, 1.0, and
1.4\AA, respectively.  Their FWHM are $\sim$13\AA, and their full widths
at zero intensity are $\sim$45\AA.  These line widths are comparable to
those of main sequence A stars but their strengths are much weaker.  The
FWHM are also similar to those of some types of hot white dwarfs (e.g.,
Feige 110), but the full width at zero intensity (i.e., the wings) of
white dwarf lines are typically wider, $\sim$75\AA.  No other stellar
absorption features are apparent either in the individual spectra or in
the co-added spectra.  The possible origin of the H absorption lines is
discussed below. 

In Figure~\ref{spec} we compare the summed spectrum of SMC 13 to that of
the prototype SSS, CAL 83.  The spectra of CAL 83 were obtained during the
same observing run with the identical instrumental configuration, so they
can be compared directly with those of SMC 13, although they have higher
S/N.  We note that the continuum of SMC 13 is much bluer than CAL 83.  (It
is also bluer than the hottest, eclipsed region in CAL 87.)  While the
emission lines are very similar in the two systems, there are some
distinct differences.  The most prominent emissions, He II 4686\AA\ and
`H$\alpha$' (we give it in quotation marks since its velocity and strength
compared to other He II lines shows it must be a blend of H and He II),
are weaker in SMC 13 than in CAL 83.  Cowley et al.\ (1997) show that SMC
13 has the weakest emission lines among six SSS they intercompare.  In SMC
13 the average equivalent width of He II-4686 is 2.9\AA\ (with FWHM =
10\AA), compared with 10\AA\ (with FWHM = 5\AA) for CAL 83.  However, the
width of 4686\AA\ is comparable to that in the eclipsing system CAL 87,
suggesting the inclination of the SMC 13 system may be high.  `H$\alpha$'
has a strongly negative velocity ($-88$ km s$^{-1}$ if it were only H).
Even if the line were entirely due to He II-6560, its mean velocity ($+36$
km s$^{-1}$) is more negative than the mean velocity of He II-4686 ($+225$
km s$^{-1}$).  This either implies some outflow from the system or a
distorted line profile perhaps due to overlying absorption.  We discuss
this further below.  O VI is weaker in SMC 13 than in CAL 83. 

The 1996 spectra are much better than, but consistent with, the ones we
obtained in 1994 (Schmidtke et al.\ 1996).  However, van Teesling et al.\
(1996) found H$\alpha$ in absorption and the other Balmer absorptions 
stronger in 1995 October than in any of our data. 

Highly shifted emission features which appear to be formed in outflows or
jets have been observed in the supersoft sources CAL 83 and RX
J0513.9$-$6951 (Crampton et al.\ 1996, Southwell et al.\ 1996).  No such
features were found in the SMC 13 spectra, but the signal-to-noise of the
spectra are much lower because of its faintness, so perhaps they would not
be detectable.

\subsubsection{Phase-related Spectral Variations}

Measurements of line strengths and velocities have been made on individual
spectra and on spectra which have been co-added in six phase bins.  These
measurements are presented in Tables 2 and 3.  Figure~\ref{coad} plots the
phase-binned spectra of SMC 13 in order of orbital phase.  Equivalent
widths and radial velocities of some lines are plotted versus phase in
Figure~\ref{lines}.  Note that although He II-4686 shows some variation of
equivalent width with phase, it is not simply a reflection of the light
curve since its maximum occurs near phase 0.8.  The `H$\alpha$' blend of
H$+$He II shows a large scatter in equivalent width, with changes between
nights being as large as any that may be phase-related. 

Figure~\ref{coad} and Table 2 show that the broad hydrogen absorptions
vary in strength and velocity through the orbital cycle.  H$\gamma$
absorption appears to be weakest near phase 0.5 when the system is
brightest.  The velocities of the absorption components have a much larger
amplitude than either He II-4686 or `H$\beta$' emission, but they are
closely co-phased with the emission.  We discuss this further below.

\subsubsection{Spatially Extended Line Emission?}

Pakull \& Angebault (1986) and Pakull \& Motch (1989) first reported
detection of spatially resolved ionized nebulae surrounding X-ray sources,
including one around the SSS CAL 83.  More recently, Rappaport et al.\
(1994) developed detailed models for ionization nebulae expected around
SSS.  Remillard, Rappaport, \& Macri (1995) then obtained [O III] and
H$\alpha$ images to search for additional such nebulae in the Magellanic
Clouds.  Although they reported successfully detecting the known nebula
around CAL 83, no nebulosity was found around SMC 13 or any of the other
eight SSS they observed. 

Examination of the average of all our long slit spectra of SMC 13
similarly shows no evidence of spatially extended line emission at [O
III], H$\beta$, or He II, although comparable observations of CAL 83 show
the extended emission at all of these lines.  Since SMC 13 is so much
fainter, we estimate that any such emission is less than 5\% of the
stellar emission lines.

\subsubsection{Radial Velocity Measurements} 

The radial velocities of the strong-line peaks were measured individually
by fitting parabolae through them and also by cross-correlating the
emission lines from individual spectra against the mean of all spectra. 
In practice, this was possible only for He II-4686 and the `H$\alpha$'
blend.  All other emission lines were too weak and noisy to derive
phase-dependent measures, even using the co-added spectra. 

He II-4686 shows a clear radial velocity variation with phase.  A circular
orbit fit to our radial velocities, using the ephemeris given in \S2.1.1,
yields a semi-amplitude K $=100\pm$19 km s$^{-1}$ with maximum positive
velocity at phase 0.75$\pm$0.03.  Given the uncertainly of both the fit to
a velocity curve and the time of minimum light, T$_0$, this phasing agrees
with what is expected if the He II velocity traces the orbital motion of
the eclipsed component (i.e. arises primarily in an accretion disk about
the compact star).  However, it is likely that the He II is not
distributed entirely symmetrically, as indicated by its equivalent width
variation which shows its maximum intensity occurs near phase 0.8. 

`H$\alpha$' may show a similarly-phased velocity variation, but with
considerably more scatter and a much lower amplitude (K$\sim$25 km
s$^{-1}$).  Since this line is clearly a blend, its He II component
probably shows the same velocity variation as 4686\AA.  But this component
is blended with a stronger H$\alpha$ component whose velocity must be
quite different from He II.  The overall negative velocity of the line
center from any possible blend of H$\alpha$ + He II, as described in
\S2.2.2, is puzzling.  We have verified in other spectra taken during the
same observing run that the wavelength scale is reliable, in spite of
being near the end of the detector.  Its negative velocity implies outflow
seen at all phases, but no receding emission is present.  Possibly the
emission line is shifted by absorption from an inflowing medium (``reverse
P-Cygni" effect) seen at all phases.  However, we know of no other such
system, and there is no obvious source of this inflowing gas.

Attempts to measure the absorption line velocities (by cross-correlation)
of H$\beta$, H$\gamma$ and H$\delta$ on the individual spectra were
unsuccessful, but measurements of phase-binned spectra show that the
absorption lines move in phase with the emission lines although with a
larger amplitude.  (See the spectra in Figure~\ref{coad} and the velocity
plot in Figure~\ref{lines}.)  Because H$\beta$ is strongly contaminated
with central emission, only H$\gamma$ was used to measure the absorption
velocities, although the variation can be seen in H$\beta$ and H$\delta$
as well. 

We have only five absorption-line velocity measures, and although their
variation does not appear very sinusoidal, there are too few points to be
certain.  A formal fit of a circular orbit to them yields a semi-amplitude
K $=$ 464$\pm$120 km s$^{-1}$ with maximum positive velocity occuring at
photometric phase 0.82$\pm$0.06, remarkably close to the He II velocity
phasing considering the small number of points used in the fit.  The mean
velocity is $\sim$200 km s$^{-1}$ lower than the systemic velocity for He
II-4686 emission (mean absorption velocity is $+21\pm113$ km s$^{-1}$
while He II emission velocity is $+217\pm14$ km s$^{-1}$).  We note that
this curiously low velocity is similar to the mean velocity found for
`H$\alpha$' emission, as described in \S2.2.2.  From the symmetry of the
H$\gamma$ absorption line, the velocities do not appear to be contaminated
by the narrow emission component, but there could be some effect.  Weak
emission contamination would cause us to overestimate this amplitude, if
the absorption has a larger velocity amplitude than the emission. 
However, it is difficult to achieve the large amplification which is
observed.

Much better spectra would be required to discuss the shape of the
absorption velocity curve in more detail.  The origin of this high
velocity amplitude is difficult to interpret.  It suggests the lines
arise from some part of the binary system centered several times further
from the center of mass than the He II emission.  It is possible that the
broad absorption arises in the optically thick accretion disk, an idea
which was first suggested for the UX UMa stars by Warner (1976) and
further developed in some detail by Mayo, Wickramasinghe, \& Whelan
(1980).  (Also see an illustration of such a profile in RZ Gru in Figure
4.6 by Warner, 1995.)  However, any disk absorption should have the same
velocity amplitude as the disk.  Eclipses by the disk edge or the
companion star can distort the descending velocity curve and even increase
its amplitude, but they cannot produce the observed large amplitude and
negative shift of the mean velocity.  On the other hand, it also appears
improbable that the absorption arises on the mass-losing star because of
its spectral signature and the phasing of the velocities.

\section{THE SMC 13 BINARY SYSTEM} 

\subsection{Orbital Velocities} 

It is clear that the period found in the photometry also fits the observed
velocities very well.  Thus, the velocities confirm that the orbital
period is $\sim$0.1719 d.  The He II velocities are properly phased for
orbital motion of the compact star.  However, in standard models for disk
hot spots, the place where the mass transfer stream impacts the disk is
the site of extra line emission, so we may not have a `clean' measure from
which to determine the stellar masses.  In SMC 13 the maximum He II line
flux occurs at phase 0.8, which is when the hot-spot region is closest to
our line of sight.  It is not possible to measure strengths of other lines
in individual spectra, but if we co-add six spectra around the He II line
maximum and six at minimum, the O VI 5290\AA\ line is stronger by a factor
two in the He II maximum bin.  Thus, some of the line emission must also
have non-orbital velocity.  However, the non-orbital velocity of material
passing though the hot-spot is likely to be small, since there are no
strong forces here and also because it occurs where the stream is stopped
by the disk.  This spot also lies near the center of mass, and thus its
velocity should not lead to an overestimate of the orbital motion.
Finally, the close agreement between spectroscopic and photometric phases
suggest that non-orbital motions are small.  We note the velocity
amplitude could be underestimated by the phase smearing of the
spectroscopic observations and by contributions from the hot-spot. 

To get some idea of the range of possible masses of the component stars,
we assume that the measured He II emission-line velocity is approximately
the same as that of the compact star's orbital motion.  The He II velocity
amplitude (K$\sim100$ km s$^{-1}$) yields a mass function, f(M)=
0.0178M$_{\odot}$.  The resulting masses for various values of the orbital
inclination are shown in Figure~\ref{mass}, where the dashed line shows
where a main-sequence star of that mass would fill its Roche lobe.  The
secondary star masses lie in the range of 0.4 - 0.5M$_{\odot}$ for cases
where this component fills its Roche lobe, regardless of the value of the
inclination.  This means that the companion star must be much fainter than
the whole system, since the apparent magnitude of a 0.5M$_{\odot}$ star at
the distance of the SMC is m$_V\sim$27.  Hence, it contributes almost
nothing to the overall luminosity of the system.  The compact-star masses
cover a wide range, depending on the orbital inclination.  For high values
of the inclination the compact star mass is near the upper limit for white
dwarfs.  We discuss below the most likely values for orbital inclination
and the implied compact-star masses. 

We caution that we cannot be certain that the measured velocities are
entirely representative of the compact star's motion.  Nevertheless, most
of the effects discussed here would result in underestimating the velocity
amplitude and hence the masses. 

\subsection{Light Curve}

\subsubsection{Models}

Schandl et al.\ (1996) and Meyer-Hofmeister et al.\ (1997) have used detailed
models to reproduce the observed light curve for the SSS CAL 87.  They
assume a relatively thick accretion disk with a high rim (which they call a
``spray") extending halfway around the disk on the ``following'' side. 
The high disk rim is caused by the impact of an accretion stream onto the
disk.  This rim is irradiated by the central white dwarf and becomes a
significant source of optical continuum light.  In their model, the thick
disk rim is opaque and decreases azimuthally away from the hot-spot.  The
thick region at the hot-spot begins to hide the central hot source before
the true eclipse by the secondary star begins.  Their analysis leads to an
orbital inclination of $i\sim78^{\circ}$ for CAL 87 when masses of 0.75
and 1.5M$_{\odot}$ are adopted for the white dwarf and the secondary star,
respectively. 

Meyer-Hofmeister et al.\ demonstrate that for other SSS they have studied,
this type of model also fits the observed light curves very well.  In the
models, light from both the disk and the X-ray heated companion star are
considered.  The disk edge may occult parts of its interior, and the
companion star's brightness depends on its aspect because of X-ray
heating.  Hence, both the disk and the star may be eclipsed by each other
or be self-occulted.  Predicted light curves for non-eclipsing
inclinations are also modelled by Meyer-Hofmeister et al.  They show that
the light curve shape is diagnostic of the inclination.  In non-eclipsing
systems (e.g., CAL 83, RX J0513.9$-$6951) the light variation is
principally due to the azimuthal variation of the reprocessed radiation
from the high disk rim with a lesser contribution from the heated inner
hemisphere of the secondary star.  In the systems they considered, the
disk is smaller than the companion star in the z-direction, so that high
inclinations are required in order for deep disk eclipses to occur.  The
light curve `maximum' is flat in an eclipsing system and curved in
non-eclipsing systems.  The light curve `minimum' is a combination of the
eclipse of the disk and viewing the unheated side of the companion star. 
The tell-tale signature of such an eclipse is the asymmetry due to the
azimuthal variation of the high disk rim.  Such a picture is consistent
with the eclipse asymmetry seen in SMC 13 (Figure~\ref{lc}). 

\subsubsection{SMC 13 Light Curve and Component Masses}

The orbital light curve of SMC 13 displays one maximum and one minimum per
cycle, similar to the behavior observed in other SSS (e.g.,
Meyer-Hofmeister et al.\ 1997), rather than showing a double-humped,
ellipsoidal variation of the companion star found in some X-ray binaries.
If the light curve of SMC 13 were due entirely to X-ray heating of the
companion star, then the light from the companion star would provide a
large fraction of the total light, and hence we might expect to see
spectral signatures from this star.  Since the only absorption lines move
in phase with He II emission, this is not possible.  (We also note that
CAL 83 and RX J0513.9$-$6951, which are considered to be low inclination
systems, have narrower emission lines, as expected from a disk viewed
nearly face-on.) Thus we conclude that a partial eclipse of a luminous
accretion disk around the compact star is the major cause of the observed
light variation in SMC 13.  The agreement in phase of the X-ray light
curve is consistent with this picture, since the X-rays must come from the
hot, compact star. 

The mean magnitude of SMC 13 is $V\sim$20.4 which corresponds to
M$_{V}\sim+1.4$, assuming a SMC distance modulus of 18.7 (van den Bergh
1992) and an average reddening E$_{B-V} =$ 0.1 mag.  For comparison, CAL
87 has a longer period (P$=$0.44 d) and is more luminous (uneclipsed
magnitude M$_{V}\sim+0.3$).  In SMC 13, the shorter period indicates a
smaller system and hence a less luminous accretion disk.  The lack of
color variation through the orbital cycle of SMC 13 suggests that the
hotter inner portions of the disk are never occulted, unlike the case in
CAL 87 where the system becomes redder during central eclipse as the
hottest regions are covered (Cowley et al.\ 1991).  Thus, we infer a
somewhat smaller inclination for SMC 13 than for CAL 87.  The SMC 13 light
curve has a flat maximum and an asymmetrical minimum, both signatures of
an eclipsing system.  The mass ratios indicated by our discussion of the
mass function set the range of possible inclinations for which partial
eclipses occur as 70$^{\circ}$ or larger.  Thus, the inclination is likely
to lie in the range $\sim70^{\circ}$ to $\sim78^{\circ}$.  This imples the
most probable mass for the compact star in SMC 13 is between 1.3 and
1.5M$_{\odot}$ (see Figure~\ref{mass}). 
 
If our measured emission-line velocities are due to orbital motion, then
the system differs from the previously studied supersoft binaries which
have more massive secondary stars.  In SMC 13 the inclination derived from
both the light curve and the emission-line widths implies a low mass
companion star and a compact-star mass very near the upper mass limit for
white dwarfs.  The system also differs from other SSS in having a lower
disk luminosity and weaker emission-line flux.  This probably arises
primarily from the small size of the accretion disk in this very short
period system, but other factors which affect the disk luminosity could
include a lower mass transfer rate or a fainter disk due to different
surface temperature or X-ray flux from the compact star. 

If the absorption-line velocities instead of the emission lines were to
indicate the orbital motion of the compact star, the resulting mass
function would be f(M)$=$1.4M$_{\odot}$, which gives impossibly large
masses (for the observed luminosity of the system) for all values of
the inclination or mass ratio.  The weakening of the absorption lines
at phase 0.5 argues against their being formed on the far side of the
disk, where the orbital lever arm is larger.  If the absorption
velocities are distorted by lower-amplitude emission, the puzzle
remains although its magnitude may be smaller than we measure.  We do
not understand the origin of these lines.

\section{Summary}

SMC 13 has the shortest period and lowest luminosity of the
spectroscopically studied SSS binaries.  (RX J0439.8$-$6809 has a slightly
shorter period (Schmidtke \& Cowley 1996) but no spectroscopic work has
been undertaken on this extremely faint system.)  SMC 13 also has the
weakest emission lines in terms of equivalent widths, and the Balmer lines
are seen in absorption unlike other SSS.  It is very blue and shows no
color change with orbital phase.  The light minimum is asymmetrical and
broad, and there are small changes in the light curve over timescales of
days to years. 

The light curve appears to be primarily due to a partial eclipse of the
disk by the mass-losing star viewed at an inclination
$i\sim75^{\circ}$.  The phasing of the He II emission velocities
suggests these lines may trace the orbital motion of the compact star.
However, the line-strength variations indicate there is a hot spot on
the disk with enhanced line emission which could lead to an
underestimate of the velocity amplitude of the compact star.  The
measured emission-line velocities and orbital inclination imply masses
near to 0.5M$_{\odot}$ for the mass donor and 1.4M$_{\odot}$ for the
compact star, if non-orbital velocites are small.  The latter value
lies near the white dwarf upper mass limit and thus appears to be
inconsistent with the `standard' SSS model in which the compact star is
low-mass white dwarf and the mass donor is a 1--2M$_{\odot}$ star.
Furthermore, the steady nuclear burning model requires a high rate of
mass flow onto the white dwarf from the companion, and it may be harder
to sustain this with a lower mass companion. 

Detailed modelling of the SMC 13 light curve should be carried out to
elaborate on the simple analysis presented here.
Higher resolution spectroscopic observations of the hydrogen absorption
lines and of `H$\alpha$' may help us understand their origin and
peculiar velocities.

\acknowledgments 
We thank the staff of CTIO for their expert assistance with the optical
observations. Tom McGrath and Jeff Crane did the analysis of the
$ROSAT$ archival X-ray data and prepared the X-ray light curve. APC and
PCS acknowledge support from NSF and NASA.


\begin{table}
\caption[]{$UBV$ Photometry of SMC 13}
\begin{tabular}{cccccc}
HJD & $U$ & error &  HJD & $U$ & error  \\
2440000+ &&& 2440000+ && \\
\hline
~9333.58743  &  19.190 & 0.061 & ~9335.60002 &  18.987 & 0.061 \\
~9665.55582  &  19.113 & 0.047 & ~9666.60808 &  19.203 & 0.057 \\
10045.55228 &  19.147 & 0.028 & 10045.60055 &  19.221 & 0.032 \\
10045.65033 &  18.967 & 0.025 & 10046.56874 &  19.121 & 0.039 \\
10049.59877 &  18.989 & 0.055 & 10389.63870 &  19.004 & 0.046 \\
10390.56703 &  18.900 & 0.035 & 10393.53105 &  19.191 & 0.027 \\
10393.57317 &  19.128 & 0.027 & 10393.61346 &  18.960 & 0.022  \\
\hline
\\[.1in]
HJD & $B$ & error &  HJD & $B$ & error  \\
2440000+ &&& 2440000+ && \\
\hline
~9330.62604 &  20.176 & 0.048 & ~9333.57590 &  20.180 & 0.039 \\
~9335.58782 &  20.120 & 0.067 & ~9665.54322 &  20.112 & 0.037 \\
~9666.59662 &  20.197 & 0.046 & 10044.57470 &  20.204 & 0.026 \\
10044.59165 &  20.163 & 0.035 & 10044.60803 &  20.069 & 0.026 \\
10044.64514 &  20.174 & 0.033 & 10044.66217 &  20.174 & 0.026 \\
10044.67831 &  20.258 & 0.035 & 10044.74061 &  20.354 & 0.037 \\
10044.75631 &  20.166 & 0.038 & 10045.54068 &  20.284 & 0.048 \\
10045.57259 &  20.318 & 0.037 & 10045.58873 &  20.383 & 0.037 \\
10045.62019 &  20.202 & 0.039 & 10045.63657 &  20.089 & 0.029 \\
10046.58133 &  20.217 & 0.026 & 10046.59797 &  20.393 & 0.035 \\
10046.60645 &  20.374 & 0.033 & 10046.61448 &  20.368 & 0.032 \\
10046.62229 &  20.382 & 0.040 & 10046.63020 &  20.291 & 0.034 \\
10046.63839 &  20.280 & 0.024 & 10046.65451 &  20.130 & 0.026 \\
10048.57762 &  20.030 & 0.047 & 10048.58613 &  20.078 & 0.062 \\
10048.59391 &  20.106 & 0.044 & 10048.60177 &  20.165 & 0.045 \\
10048.60964 &  20.035 & 0.053 & 10049.58680 &  20.079 & 0.050 \\
10389.62732 &  20.181 & 0.033 & 10389.65323 &  20.086 & 0.024 \\
10390.55565 &  20.040 & 0.029 & 10390.59916 &  20.085 & 0.028 \\
10390.61612 &  20.210 & 0.038 & 10390.63383 &  20.228 & 0.039 \\
10390.65206 &  20.219 & 0.032 & 10390.66888 &  20.105 & 0.037 \\
10393.51962 &  20.196 & 0.035 & 10393.54295 &  20.395 & 0.039 \\
10393.55883 &  20.359 & 0.042 & 10393.58530 &  20.221 & 0.041 \\
10393.60173 &  20.071 & 0.027 & 10394.59813 &  20.346 & 0.034 \\
10394.61495 &  20.084 & 0.026 \\
\hline

\end{tabular}
	
\end{table}

\begin{table}
\begin{tabular}{cccccc}
HJD & $V$ & error &  HJD & $V$ & error  \\
2440000+ &&& 2440000+ && \\
\hline
9330.61758 & 20.231 & 0.037 & 9332.54302 & 20.430 & 0.048 \\  
9333.56800 &  20.343 & 0.045 & 9334.55199 &  20.329 & 0.041 \\ 
9334.60883 &  20.347 & 0.039 & 9334.66236 &  20.531 & 0.051 \\ 
9335.54130 &  20.464 & 0.054 & 9335.54936 &  20.481 & 0.050 \\ 
9335.55707 &  20.492 & 0.070 & 9335.56453 &  20.363 & 0.052 \\ 
9335.57229 &  20.404 & 0.067 & 9335.57972 &  20.349 & 0.065 \\ 
9659.52743 &  20.203 & 0.026 & 9659.53779 &  20.263 & 0.025 \\ 
9659.54579 &  20.294 & 0.034 & 9659.55367 &  20.340 & 0.039 \\ 
9659.56155 &  20.333 & 0.025 & 9659.57129 &  20.440 & 0.025 \\
9659.57917 &  20.496 & 0.034 & 9659.58705 &  20.435 & 0.027 \\
9659.59492 &  20.477 & 0.033 & 9659.60278 &  20.499 & 0.036 \\
9659.61065 &  20.489 & 0.036 & 9659.61851 &  20.480 & 0.039 \\
9659.62678 &  20.392 & 0.032 & 9659.63516 &  20.347 & 0.027 \\
9659.64301 &  20.313 & 0.026 & 9659.65087 &  20.371 & 0.045 \\
9659.65888 &  20.308 & 0.027 & 9659.66677 &  20.323 & 0.027 \\
9659.74022 &  20.395 & 0.032 & 9659.74920 &  20.466 & 0.035 \\
9659.75708 &  20.472 & 0.034 & 9660.51450 &  20.256 & 0.033 \\
9660.52322 &  20.258 & 0.029 & 9660.53108 &  20.202 & 0.033 \\
9660.53899 &  20.245 & 0.024 & 9660.54686 &  20.207 & 0.026 \\
9660.55482 &  20.240 & 0.027 & 9660.56272 &  20.251 & 0.027 \\
9660.57062 &  20.273 & 0.031 & 9660.57864 &  20.185 & 0.030 \\
9660.58654 &  20.286 & 0.027 & 9660.59439 &  20.362 & 0.031 \\
9660.60226 &  20.392 & 0.024 & 9660.61016 &  20.456 & 0.025 \\
9660.61802 &  20.403 & 0.029 & 9660.62649 &  20.464 & 0.033 \\
9660.63517 &  20.573 & 0.039 & 9660.64363 &  20.492 & 0.027 \\
9660.65149 &  20.414 & 0.033 & 9660.65941 &  20.424 & 0.034 \\
9660.66745 &  20.341 & 0.032 & 9660.67529 &  20.269 & 0.023 \\
9660.68312 &  20.257 & 0.027 & 9660.74863 &  20.360 & 0.036 \\
9660.75729 &  20.362 & 0.054 & 9660.76518 &  20.458 & 0.086 \\
9662.56354 &  20.266 & 0.025 & 9662.60377 &  20.206 & 0.025 \\
9662.64551 &  20.249 & 0.030 & 9664.56456 &  20.438 & 0.046 \\
9664.58180 &  20.540 & 0.047 & 9664.59015 &  20.614 & 0.054 \\
9664.59804 &  20.482 & 0.034 & 9664.60596 &  20.486 & 0.043 \\
9664.61386 &  20.413 & 0.037 & 9664.62195 &  20.365 & 0.047 \\
\hline

\end{tabular}
	
\end{table}

\begin{table}
\begin{tabular}{cccccc}
HJD & $V$ & error &  HJD & $V$ & error  \\
2440000+ &&& 2440000+ && \\
\hline
~9665.51520 &  20.294 & 0.075 & ~9665.53453 &  20.284 & 0.054 \\
~9665.56831 &  20.398 & 0.045 & ~9665.57624 &  20.384 & 0.045 \\
~9665.58429 &  20.360 & 0.053 & ~9665.59205 &  20.401 & 0.039 \\
~9665.59986 &  20.450 & 0.047 & ~9665.60768 &  20.560 & 0.052 \\
~9665.61559 &  20.532 & 0.038 & ~9665.62344 &  20.551 & 0.041 \\
~9665.63111 &  20.480 & 0.033 & ~9665.63873 &  20.436 & 0.047 \\
~9666.58841 &  20.297 & 0.046 & 10044.56651 &  20.460 & 0.034 \\
10044.58320 &  20.313 & 0.029 & 10044.59998 &  20.313 & 0.046 \\
10044.63706 &  20.256 & 0.026 & 10044.65368 &  20.260 & 0.022 \\
10044.67029 &  20.355 & 0.029 & 10044.73204 &  20.449 & 0.029 \\
10044.74846 &  20.417 & 0.042 & 10045.53240 &  20.388 & 0.030 \\
10045.56471 &  20.455 & 0.035 & 10045.58060 &  20.492 & 0.032 \\
10045.61227 &  20.339 & 0.024 & 10045.62860 &  20.296 & 0.027 \\
10046.58981 &  20.431 & 0.026 & 10046.64669 &  20.322 & 0.032 \\
10049.57870 &  20.217 & 0.035 & 10389.61874 &  20.390 & 0.044 \\
10389.66136 &  20.199 & 0.029 & 10390.54777 &  20.164 & 0.032 \\
10390.59079 &  20.218 & 0.027 & 10390.60755 &  20.256 & 0.027 \\
10390.62530 &  20.370 & 0.032 & 10390.64218 &  20.436 & 0.028 \\
10390.66051 &  20.278 & 0.029 & 10390.67679 &  20.207 & 0.033 \\
10391.52999 &  20.321 & 0.041 & 10391.58165 &  20.211 & 0.025 \\
10391.62174 &  20.221 & 0.028 & 10392.57174 &  20.235 & 0.032 \\
10392.63801 &  20.210 & 0.028 & 10393.51140 &  20.267 & 0.036 \\
10393.55092 &  20.478 & 0.035 & 10393.59355 &  20.264 & 0.028 \\
10394.60680 &  20.367 & 0.033 & 10394.62378 &  20.240 & 0.025 \\
\hline

\end{tabular}
	
\end{table}

\clearpage

\centerline{CAPTIONS TO FIGURES}

\figcaption[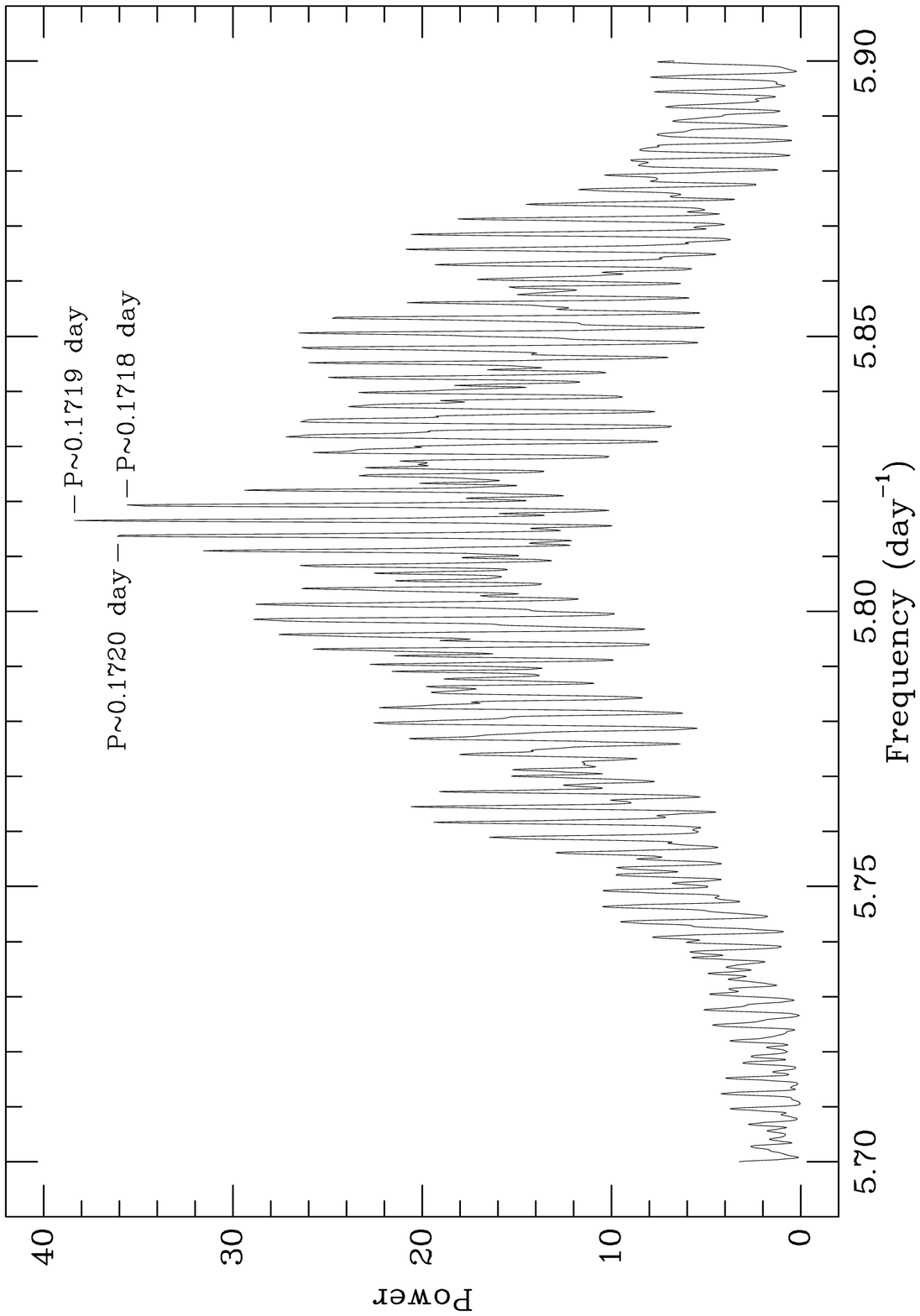]{Periodogram for SMC 13 based on $V$ photometry
obtained in 1993, 1994, 1995, and 1996. \label{per}}

\figcaption[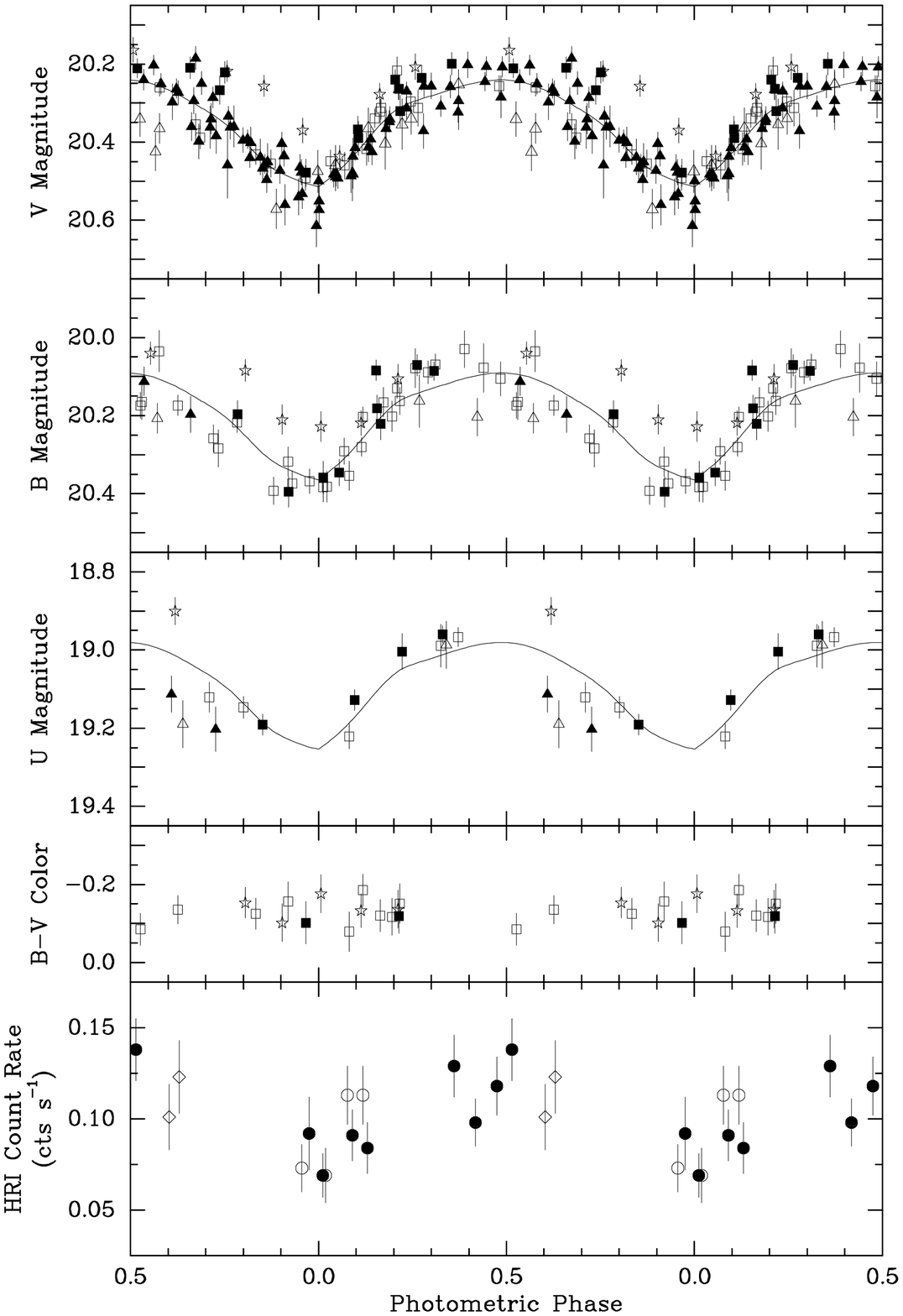]{$UBV$-filter light curves and $B-V$ photometry of SMC
13 from data obtained during observing runs in 1993 (open triangles), 1994
(filled triangles), 1995 (open squares), 1996 (excluding night 2, filled
squares), and 1996 night 2 (open stars).  The ephemeris given in the text
is used.  Also shown plotted on the same period are the $ROSAT$-HRI X-ray
data obtained from the archives. \label{lc}} 

\figcaption[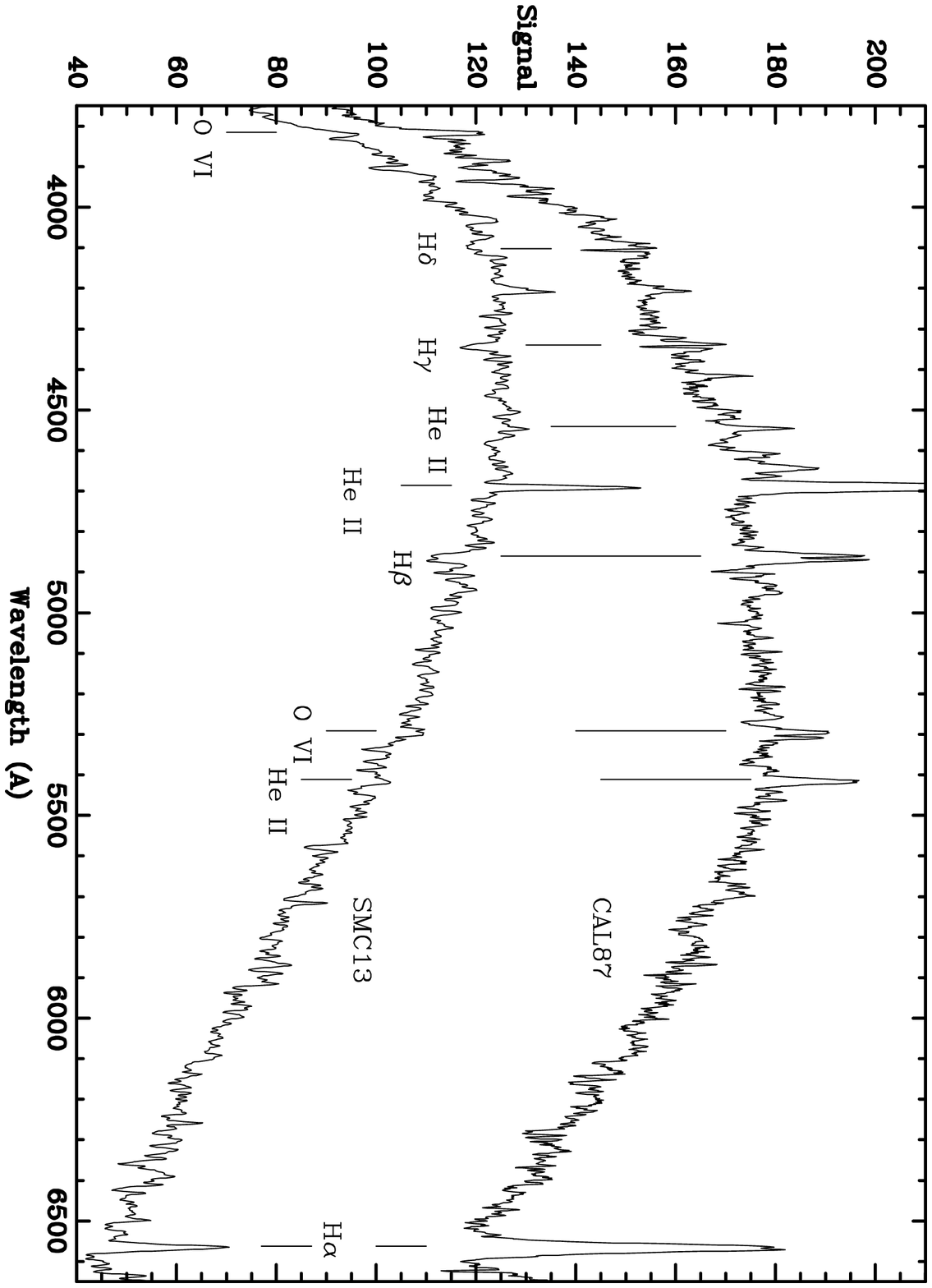]{Mean spectrum of SMC 13 compared with mean of CAL 83
from the same observing run.  Principal lines are marked. \label{spec}} 

\figcaption[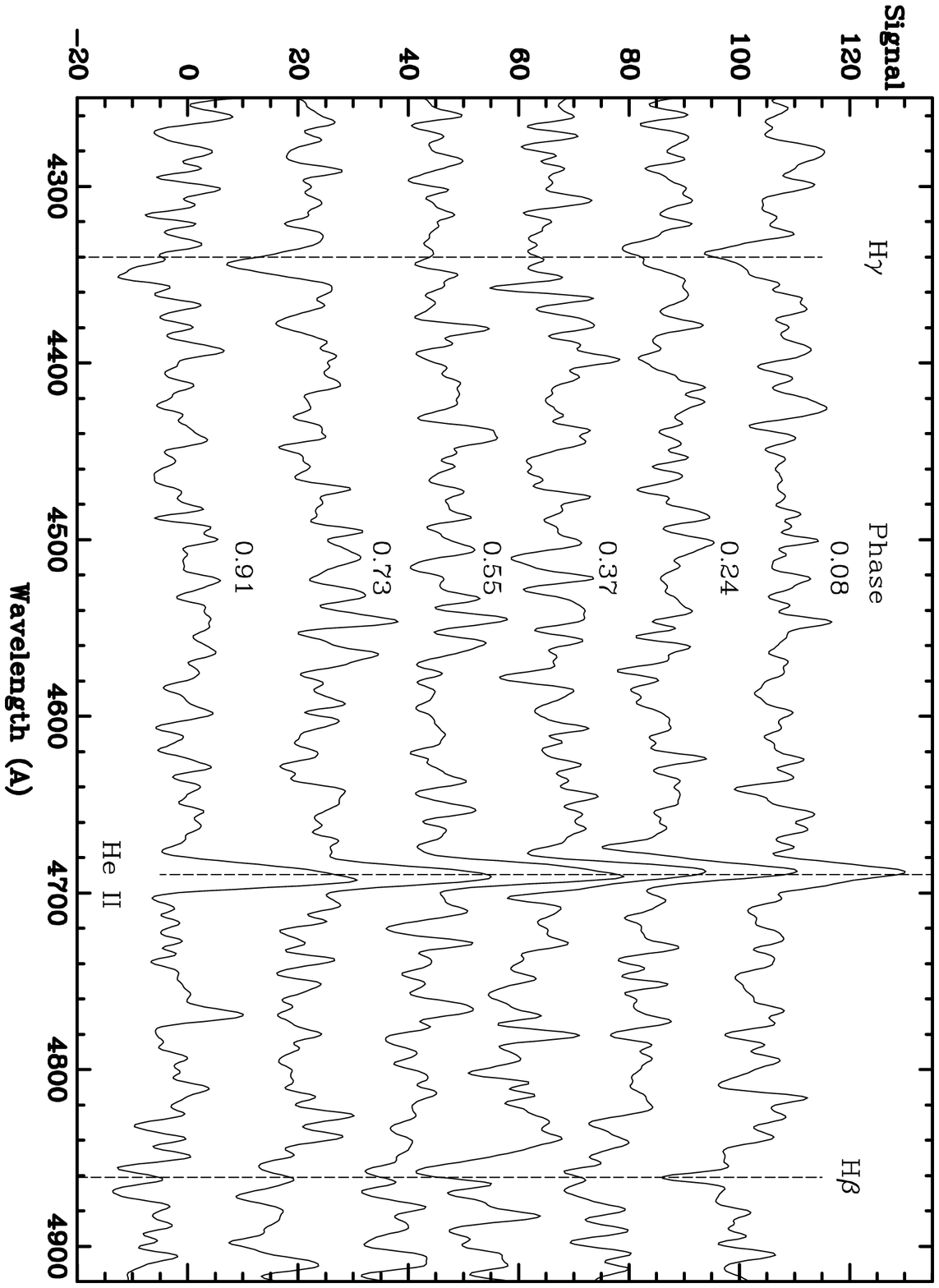]{Co-added spectra in six phase bins, showing Balmer-line
velocity changes. \label{coad}}

\figcaption[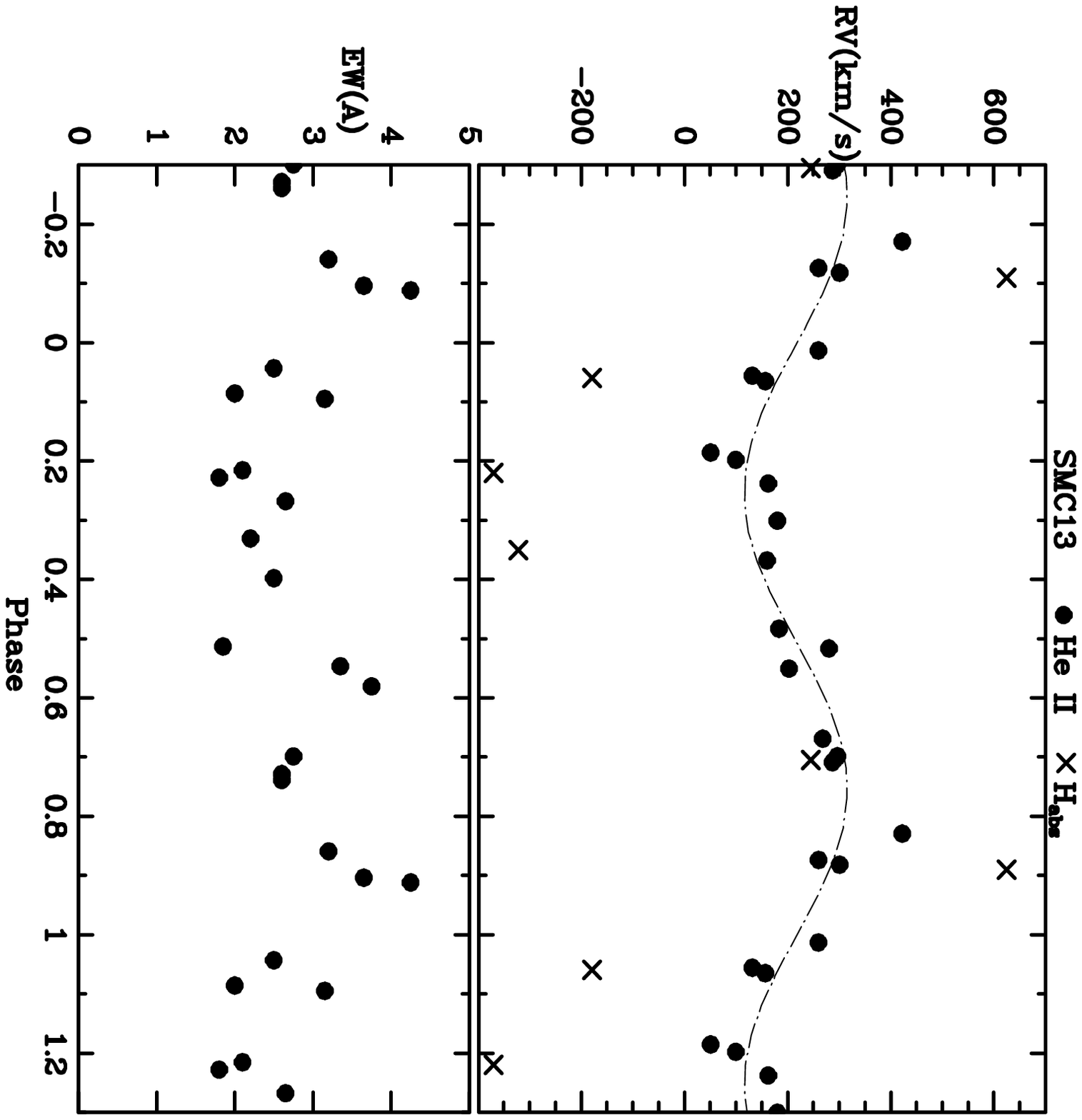]{Phase variation of radial velocities and equivalent
widths of He II-4686 emission (filled circles) and H absorption (crosses).
The dashed curve is the fit of a circular orbit to the emission-line
velocities. \label{lines}} 

\figcaption[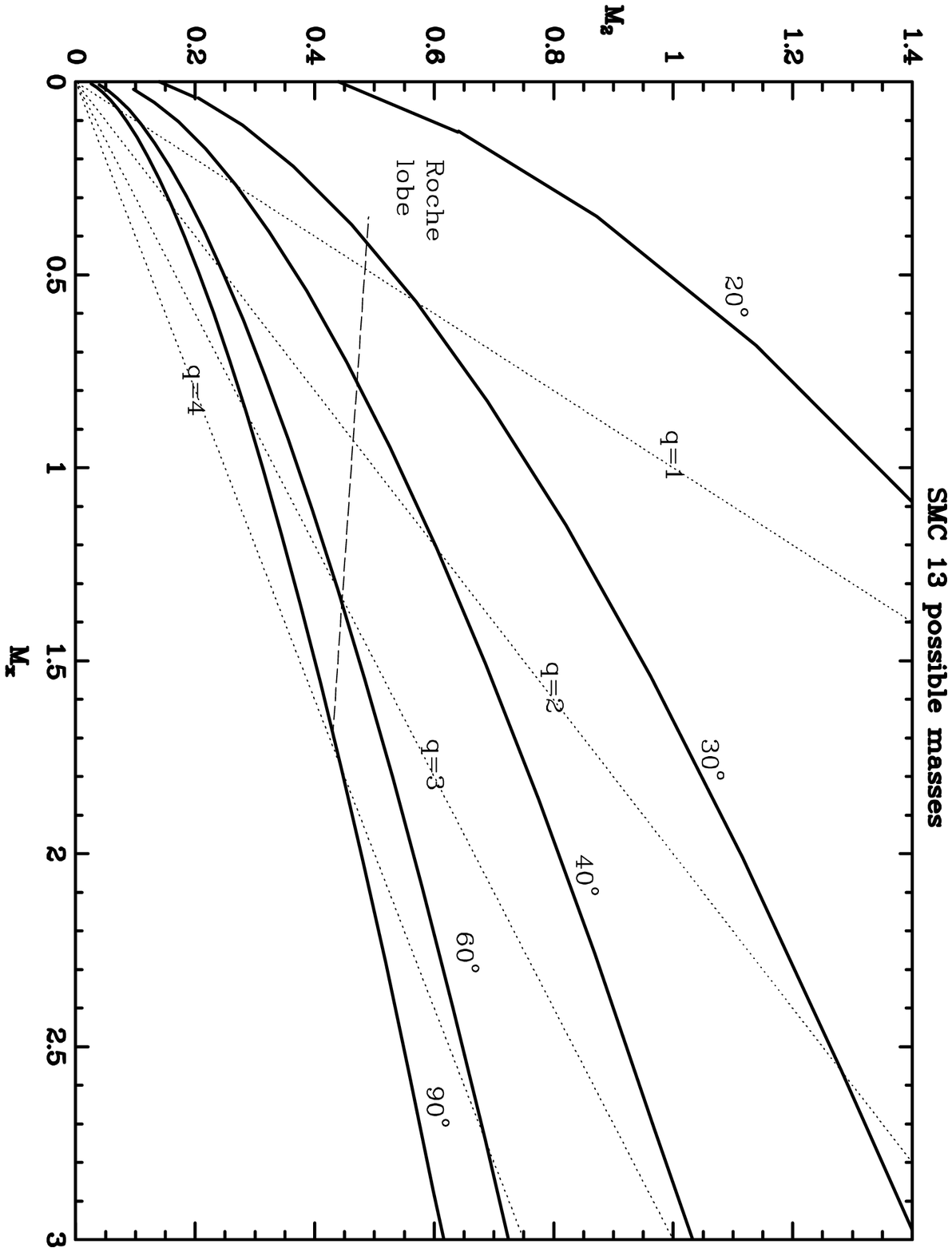]{SMC 13 mass diagram for He II emission velocities.  The
dashed line shows where a main-sequence star of mass M$_2$ fills its Roche
lobe.  Most probable masses lie on or just below this line for inclination
values near $i\sim75^{\circ}$. \label{mass}}

\plotone{smc13fig1.ps}

\plotone{smc13fig2.ps}

\plotone{smc13fig3.ps}

\plotone{smc13fig4.ps}

\plotone{smc13fig5.ps}

\plotone{smc13fig6.ps}

\end{document}